\documentclass{aa}

\usepackage[breaklinks, colorlinks, citecolor=blue, urlcolor = blue]{hyperref}
\usepackage{comment}

\makeatletter
\renewcommand*\aa@pageof{, page \thepage{} of \pageref*{LastPage}}
\makeatother

\def\xmm{{\it XMM-Newton}}
\def\suzaku{{\it Suzaku}}
\def\nustar{{\it NuSTAR}}

\usepackage{graphicx}
\usepackage{orcidlink}

\usepackage{txfonts}
\usepackage{hyperref}
\begin{document}

\title{Energy dependence of the X-ray power spectrum in NGC 4051 and NGC 4395}


\author{V. A. Diamantopoulos \inst{\ref{uoc1}}\orcidlink{0009-0005-7899-3705}
\and
I. E. Papadakis \inst{\ref{uoc2},\ref{ia}}\orcidlink{0000-0001-6264-140X}
\and
A. Akylas \inst{\ref{noa}}\orcidlink{0000-0002-3214-3589}
\and
A. Zoghbi \inst{\ref{maryland},\ref{heasarc},\ref{GSFC1}}\orcidlink{0000-0002-0572-9613}
\and
E. Kammoun \inst{\ref{Caltech}}\orcidlink{0000-0002-0273-218X}
\and
B. Rani \inst{\ref{GSFC2},\ref{CSST-UMary}}\orcidlink{0000-0001-5711-084X}
}
\institute{
Department of Physics, University of Crete, 71003 Heraklion, Greece, \email{\href{mailto:vasdiam@physics.uoc.gr}{vasdiam@physics.uoc.gr}}\label{uoc1}
\and
Department of Physics and Institute of Theoretical and Computational Physics, University of Crete, 71003 Heraklion, Greece \label{uoc2} 
\and 
Institute of Astrophysics, FORTH, GR-71110 Heraklion, Greece\label{ia} 
\and
IAASARS, National Observatory of Athens, Ioannou Metaxa and Vasileos Pavlou 15236 Athens, Greece \label{noa}
\and
Department of Astronomy, The University of Maryland, College Park, MD 20742, USA \label{maryland}
\and
HEASARC, Code 6601, NASA/GSFC, Greenbelt, MD 20771, USA \label{heasarc}
\and
CRESST II, NASA Goddard Space Flight Center, Greenbelt, MD 20771, USA \label{GSFC1}
\and
Cahill Center for Astrophysics, California Institute of Technology, 1216 East California Boulevard, Pasadena, CA 91125, USA \label{Caltech}
\and
NASA Goddard Space Flight Center, Greenbelt, MD 20771, USA \label{GSFC2}
\and
Center for Space Science and Technology, University of Maryland Baltimore County, 1000 Hilltop Circle, Baltimore, MD 21250, USA \label{CSST-UMary}
}

   \date{Received 8 October 2025 / Accepted 29 January 2026}

 
  \abstract
   {Active galactic nuclei (AGNs) exhibit significant variability across the electromagnetic spectrum on a wide range of timescales. This variability is particularly extreme in the X--ray band, showing both high-amplitude and rapid fluctuations. Power spectral density (PSD) analysis is a common tool used to characterise the observed variability in these objects. Although PSDs of AGN are typically well described by a bending power-law model, the dependence of the model parameters on photon energy has not been systematically explored in the past.}
   {Our objective is to investigate whether the PSD parameters of AGNs depend on energy, considering two low-mass Seyfert galaxies -- NGC 4051 and NGC 4395 -- as case studies. Both sources are highly variable in X-rays, and they have been observed extensively with current and past X-ray satellites. Using all archival observations, we aim to measure their power spectra over a broad frequency and energy range and investigate if and how the power spectrum evolves with energy.}
   {We used archival data from \xmm, \suzaku, and \nustar, and computed the power spectrum in six energy bands from 0.3 to 20 keV. Then we fitted the power spectra with a bending power-law model and investigated the energy dependence of the model parameters.}
   {The power spectra computed using light curves taken from different satellites and at different times are consistent within the errors, indicating that the X-ray variability process is stationary in these two objects. We found that, for both sources: a) the PSD bending frequency remains constant with the energy, b) the high-frequency slope becomes flatter with increasing energy, and c) the power spectrum amplitude decreases with increasing energy.}
   {Our results can significantly constrain current models that explain the variability of X-rays in AGNs (such as the fluctuating accretion rate model). Similar studies of more AGNs are necessary to quantify in detail the energy dependence of the power spectrum in AGNs.}

   \keywords{Active Galactic Nuclei --
                X-ray Variability --
                PSD Analysis -- NGC 4051 -- NGC 4395
               }
   \maketitle
%


\section{Introduction}\label{intro}

Active galactic nuclei (AGNs) display significant variability across the electromagnetic spectrum. The most exceptional variability, both in terms of timescales and amplitude, is exhibited in the X-rays. It is generally believed that the X-ray variations originate from the innermost region around the supermassive black hole (SMBH) and can provide insights into the accretion processes and physical conditions that drive AGN emission.

A bending power-law (BPL) model describes well the X-ray power spectral density (PSD) of Seyfert galaxies in the 2--10 keV band. Their PSDs have a power law (PL)-like shape at low frequencies (with a slope of $\sim -1$) that 'bends' or 'breaks' to a steeper slope above a characteristic frequency, $\nu_b$ \cite[see][and references therein] {paolillo-papadakis2025}. The bending frequency appears to scale with the physical parameters of the AGN. In particular, higher break frequencies (shorter characteristic timescales) are observed in AGNs of a lower black hole (BH) mass and a higher accretion rate (\citealt{mchardy-etal2006}). 

The majority of X-ray PSD studies so far have relied on the use of broadband X-ray light curves, mainly in the 2-10 keV band. Although this approach maximises the signal-to-noise ratio of the PSD, it obscures any energy-dependent variations in the PSD characteristics.

It was clear from the first detailed PSD studies of AGNs that the power spectrum varies with energy. For example, \citet{papadakis-lawrence1995} found that the slope of the X-ray power spectrum of NGC 4051 flattens at high energies, which has significant implications regarding the nature of the X-ray emitting source in AGNs. The flattening of the high-frequency slope of the PSD has been verified many times in the past. In particular, the study of the X-ray PSDs of many AGNs by \citet{gonzalez-martin-vaughan2012} showed that the slope of the X-ray PSDs in the 0.2--2 keV band is systematically steeper than the PSD slope in the 2--10 keV band. 

However, the dependence of the PSD characteristics on energy, in particular the energy dependence of the bending frequency and of the PSD amplitude, is not well known. \citet{vaughan-etal2003}, \citet{markowitz-etal2007}, \citet{markowitz2009}, and \citet{markowitz2010} studied the PSDs of MCG-6-30-15, Mrk 766, IC 4329A, and NGC 7469 in various energy bands. They reported the PSD characteristics in the various energy bands, but they did not study in detail their dependence on energy. Recently, \citet{rani-etal2025} presented a study of the NGC 4051 X-ray power spectrum using observations from \xmm\ and \textit{NICER}. 
Their main objective was to study the energy dependence of the power spectrum, by computing the PSD in many bands in the 0.3--3 keV range. This range is rather limited (by the sensitivity of {\it NICER}), although it spans an order of magnitude in energy. They found no significant energy dependence of any of the PSD characteristics in these energy bands. This result is interesting, but it is based on the study of the soft X-ray band variations only, and leaves the study of the variations at higher energies unexplored. 

We study in detail the energy dependence of the power spectrum of NGC 4051 and NGC 4395. We chose these sources because they have been extensively observed by all recent X-ray satellites and they are relatively bright (especially NGC 4051) and highly variable. Therefore, it is possible to accurately determine their X-ray PSD in many energy bands and study its energy dependence. The mass of their central BH is also small; therefore, the expected bending frequency will be well determined by \xmm\ light curves. This greatly facilitates the accurate determination of the PSD amplitude and its high-frequency slope. 

NGC 4051 is a narrow-line Seyfert 1 galaxy with a relatively low-mass BH ($ 9\pm\times 10^5 M_{\odot}$\footnote{BH mass estimates for both sources are taken from 'The AGN Black Hole Mass Database' \citep{agndatabase}, assuming a scale factor $f=4.8$, following \cite{Batiste17}.}). Since it is one of the brightest and most variable AGNs in X-rays in the local Universe, NGC 4051 has been observed by many X-ray satellites and its X-ray power spectrum has been studied extensively in the past (e.g. \citealt{lawrence-etal1987}; \citealt{papadakis-lawrence1995}; \citealt{mchardy-etal2004}; \citealt{vaughan-etal2011}), providing an abundance of data. 

\cite{mchardy-etal2004} studied in detail the X-ray PSD of the source using the long-term {\it RXTE} and a single \xmm\ observation. The resulting PSD in the 2--10 keV band is one of the best X-ray PSDs among AGN. It extends over almost eight orders of magnitude in frequency, from very low to high frequencies. It is one of the first Seyferts whose PSD was shown to be well fitted by a BPL. The PSD slope at low frequencies is very close to $\sim -1$, and then slowly bends to a slope of $\sim -2$ above $\sim 8\times 10^{-4}$ Hz. 

The PSD shape is remarkably similar to that of Cyg X-1 in its soft state, after accounting for the $\sim 10^5$ difference in the BH mass between the two sources. \cite{mchardy-etal2004} and \cite{rani-etal2025} studied the energy dependence of the power spectrum of NGC 4051. Compared to these works, we use more data and compute the PSD at longer and shorter frequencies and over a wider range of energies, which extend from $\sim 0.4$ up to 15 keV. 

NGC 4395 is a dwarf spiral galaxy. Its central active nucleus (classified as Type 1.8 Seyfert) displays even more extreme variability on shorter timescales because it hosts one of the lowest-mass BHs among all known Seyferts ($\sim 3\times 10^5 M_{\odot}$). This implies strong variations on very short timescales in its inner region, where X-rays should be produced. This was confirmed by the first long \xmm\ observation of NGC 4395. \citet{vaughan-etal2005} studied the energy dependence of the power spectrum. As with NGC 4051, we use more data and we compute the power spectra at longer and shorter frequencies and over a wider energy range. 

In this work, we present the results from the systematic analysis of the energy-dependent X-ray variability of NGC 4051 and NGC 4395. We used archival data from three different observatories, namely \xmm, \suzaku, and \nustar, which provide broad energy coverage and long continuous light curves. The combined use of the light curves from these observatories enabled us to accurately estimate the power spectrum from 0.4 to 15 keV (that is, a factor of 40 difference from the lowest to the highest energy). In this way, we can investigate how the PSD parameters (i.e. slope, amplitude, and bending frequency) change with respect to the emitted photon energy. 

The paper is structured as follows. In Sect.\,\ref{Obs-Data-Red} we discuss the observations and data analysis. In Sects.\,\ref{PSD-est} and \ref{modelfits} we discuss how we calculated the PSDs in the various energy bands and the model fits, respectively. In Sect.\,\ref{energy-dep} we present the results of the study of the energy dependence of the PSD parameters.  Finally, in Sect.\,\ref{disc} we explore the implications of our findings. 


\section{Observations and data reduction}\label{Obs-Data-Red}

We used observations from \xmm, \nustar, and \suzaku\ X-ray satellites. The combination of light curves from these X-ray missions enabled the construction of PSDs in a wide range of frequencies across seven energy bands, namely 0.3--0.5, 0.5--0.8, 0.8--1.5, 1.5--5, 5--10, 10--15, and 15--20 keV. The observation log, including the observation IDs, start time, and duration of the observations, is shown in Table \ref{table:obs_log}. We used \xmm\ EPIC-pn data due to its large effective area, which produces high signal-to-noise light curves. The long duration of the \suzaku\ and \nustar\ light curves allowed us to compute the power spectrum at frequencies lower than the lowest frequency we can probe with \xmm\ EPIC-pn light curves. We considered all the publicly available archival observations until the end of 2024 in the science archive of the three satellites for both sources. This amounted to nineteen \xmm\ observations, three from \suzaku, and five from \nustar\ for NGC 4051. There are 17 archival \xmm\ observations for NGC 4395, but only 11 of them are listed in Table \ref{table:obs_log}. Five of the remaining observations were heavily affected by background flares, and the source is very faint in the last one. Furthermore, we also considered the sole \suzaku\ observation and all archival \nustar\ observations for this source.

\subsection{Data reduction of the \xmm\ observations}
\label{Red-XMM}

We reduced the \xmm\ data using the Science Analysis System (SAS) version 21.0.0. We first filtered the data to retain only single and double pixel events (PATTERN$\leq$4), and applied the standard FLAG=0 selection expression. In order to extract light curves from the filtered data, we considered a circular source region with a radius of 40 arcseconds.

All NGC 4051 observations were conducted in the small window mode (SWM), except one (ObsID 0157560101) which was in the large window mode (LWM). For the SWM mode observations we chose rectangular background regions, which were approximately 1.15 times the size of the source aperture. The background region for the LWM observation was chosen in the same CCD as the source and was approximately 8.75 times the area of the source extraction region. The NGC 4395 observations were conducted in the full frame mode (FFM), with the exception of ObsID 0744010101 which was in the SWM. The ratio of the area of the background region versus the source region for the FFM observations was approximately 18.4, while the background region of the SWM observation was approximately 5.1 times larger than the source extraction region.

Before the extraction of light curves, we checked all observations for the presence of pile-up. All were found to be pile-up-free, except for the LWM observation of NGC 4051. To reduce the effects of pile-up, we extracted the source light curve from an annular region with an inner radius of 17.5 arcseconds and an outer radius of 40 arcseconds. This adjustment resulted in a ratio of background versus source extraction region of approximately 10.8. We also investigated the possibility that observed variations are contaminated by flaring background. To this end, we examine the background time series in each energy band. Most of the time and for all observations, the background is low and relatively stable. However, the background rate increased significantly in the final (or first) few kiloseconds of some observations. We do not consider these light curve parts in our analysis (the 'duration' listed in Table \ref{table:obs_log} indicates the duration of the light curves after we omitted the parts with high background flares). 

We extracted background-subtracted light curves using \texttt{epiclccorr}, in the energy bands between 0.3--10 keV. The \xmm\ light curves were binned at a 10 s resolution for all energy bands and for both sources. These light curves will be used for the estimation of the high-frequency part of the power spectrum of the sources. 


\subsection{Data reduction of the \suzaku\ observations}\label{Red-Su}

We reduced the \suzaku\ data using \textit{HEASoft} version 6.34 with the latest available calibration files. The light curves were extracted across the same energy bands as \xmm, using \texttt{xselect} to filter events from the source and background regions. The background light curves were scaled to match the source region size and subsequently subtracted from the corresponding source light curves. 

We extracted light curves binned at a 10 s resolution for both NGC 4051 and NGC 4395. We then rebinned them at a time resolution of 2900 s, which is approximately half the orbital period of \suzaku. This is necessary to construct evenly sampled light curves with fully exposed bins (see below). The \suzaku\ light curves with a bin size of 2900 s will be used to compute the power spectrum at low frequencies. 

Once the background-subtracted light curves were extracted, we examined whether we could add the XIS0, XIS1, and XIS3 light curves. The XIS0 and XIS3 units on the \suzaku\ satellite contained identical CCD detectors. However, XIS1 was equipped with a back-illuminated CCD, which had a different response compared to the CCDs of XIS0 and XIS3. To determine whether we could add all light curves together, we subtracted the count rates for each pair of detectors and performed a $\chi^2$-test to investigate whether the rate differences are constant. For the NGC 4051 observations, we found that the rate difference between the XIS1 and the other two light curves is not constant for most observations, in all bands (the probability of the null hypothesis of a constant difference was less than 0.01). On the other hand, we found that the difference between the XIS0 and XIS3 light curves displayed a constant difference. Therefore, we added the XIS0 and XIS3  light curves together. 

The same test was performed for the NGC 4395 light curves from the single \suzaku\ observation. We found that the difference between the count rate of the three units had no significant deviation from a constant for all pairs of detectors in the three higher energy bands. In contrast, the differences between the XIS0 and XIS1, and the XIS1 and XIS3 light curves showed a significant deviation from a constant value ($p_{null} \ll 0.01$) in the 0.3--0.5, and 0.5--0.8 keV energy bands. Therefore, we added only the XIS0 and XIS3 light curves in the 0.3--0.5, and 0.5--0.8 keV energy bands, while we added all three light curves in the other bands. 


\subsection{Data reduction of the \nustar\ observations}\label{Red-Nu}

The \nustar\ data were processed using the NuSTAR data analysis pipeline, \texttt{nupipeline}, to produce the event files used for the extraction of the light curves. For this process, we used the \texttt{nustardas v2.0.0} software package and CALDB version 20170222. From the cleaned event files, we extracted source and background light curves for the two focal plane modules, FPMA and FPMB, using the \texttt{nuproducts} script. A circular source region with a radius of 60 arcseconds was chosen for both detectors, and the standard live-time, point spread function, and vignetting corrections were applied. Background light curves were obtained from a source-free region with a radius of 120 arcseconds, located at a similar off-axis angle to the source. 

We extracted light curves binned at a 10 s resolution both for NGC 4051 and NGC 4395 in the 5--10 and 10--20 keV bands. The resulting \nustar\ light curves were then rebinned at $\Delta t=$ 2900 s, as we did with the \suzaku\ observations. We used the 10-sec binned light curves to estimate the high-frequency PSD in the 5--10 and 10--20 keV bands and the 2900 s to estimate the low-frequency PSD in the same bands. 

We performed the same test as we did for the \suzaku\ light curves to decide whether we could combine the FPMA and FPMB light curves. We found that the count rate difference of the two detectors is constant with time in all cases; thus, we added the FPMA and FPMB light curves together (for both sources). 

\begin{figure}[!h]
\centering
\includegraphics[width=0.5\textwidth]{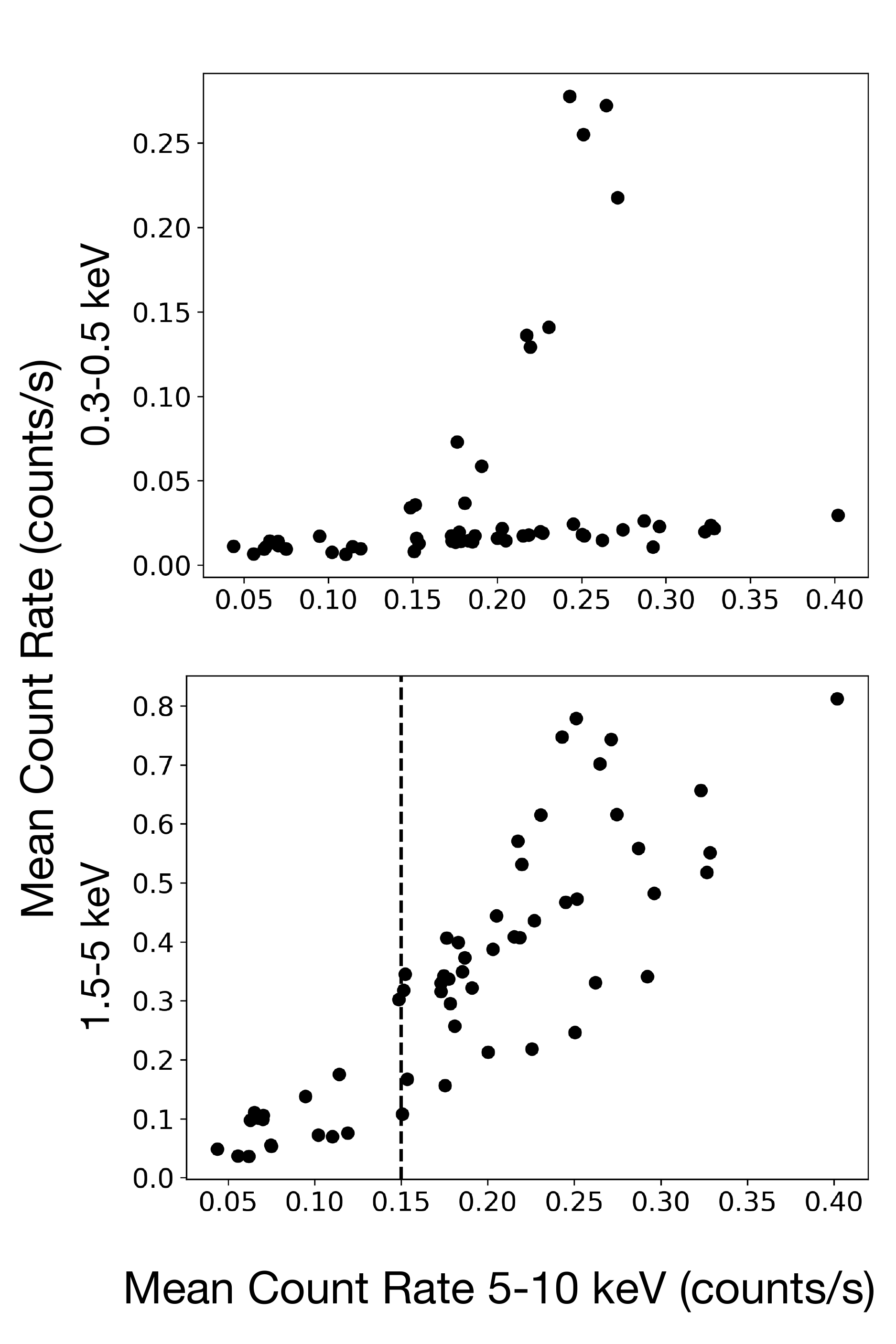}
\caption{Plot of the 0.3--0.5 and of the 1.5-5 keV keV band count rates as function of the 5--10 keV band count rate of NGC 4395 (top and bottom panels, respectively).}
\label{fig:countrates}
\end{figure}

\section{Power spectrum estimation}\label{PSD-est}

\subsection{Estimation of the high-frequency PSD}

To estimate the power spectrum at high frequencies, we used the 10-sec binned \xmm\ light curves in the 0.3--10 keV bands and the 10-sec binned \nustar\ light curves in the 5--20 keV bands.  First, we divided the light curves into shorter segments and calculated the periodogram for each one as follows:
\begin{equation}\label{eq:periodogram}
    I_N(f_j)=\frac{2 \Delta t}{N} \left\lvert \sum_{i=1}^{N} x_n(t_i) e^{-2\pi i f_j t_i} \right\rvert^2,
\end{equation}
where $\Delta t$ is the time bin of the light curve, and $N$ is the total number of points in the segment. Data were normalised to the light curve mean, i.e. $x_n(t_i)= (x(t_i)-\overline x)/\overline x$, where $\overline x$ is the mean count rate of the segment. We computed the periodogram at the usual Fourier set of frequencies, i.e. $f_j=j/(N \Delta t), \ j=1,2,...,N/2$.

We decided to work with the logarithm of the periodogram (whose distribution is much more symmetric than the distribution of $I_N$). We therefore calculated the logarithm of the periodogram; that is, log [${\rm PSD}(f_j)]=\log[I_N(f_j)]+0.25068$\footnote{The value of the constant is taken from \cite{vaughan2005}, and it is necessary to account for the fact that $I_N$ is distributed as a $\chi^2$ variable with two degrees of freedom \citep[for details see][]{papadakis-lawrence1993}.}. Our final estimate of the logarithm of the power spectrum is equal to the mean of the log-periodograms of all segments, i.e.

\begin{equation} 
\log[{\rm PSD}_{final}(f_j)]=\frac{1}{n_{seg}}\sum_{i=1}^{n_{seg}} \log[{\rm PSD}_i(f_j)],
\end{equation}

\noindent where $n_{seg}$ is the number of light curve segments in each band. The error of the average log periodogram is known and is equal to $\sigma_{PSD_{final}}(f_j)=\sqrt{0.31/n_{seg}}.$

\begin{figure*}[htbp]
  \centering
  \includegraphics[width=0.9\textwidth]{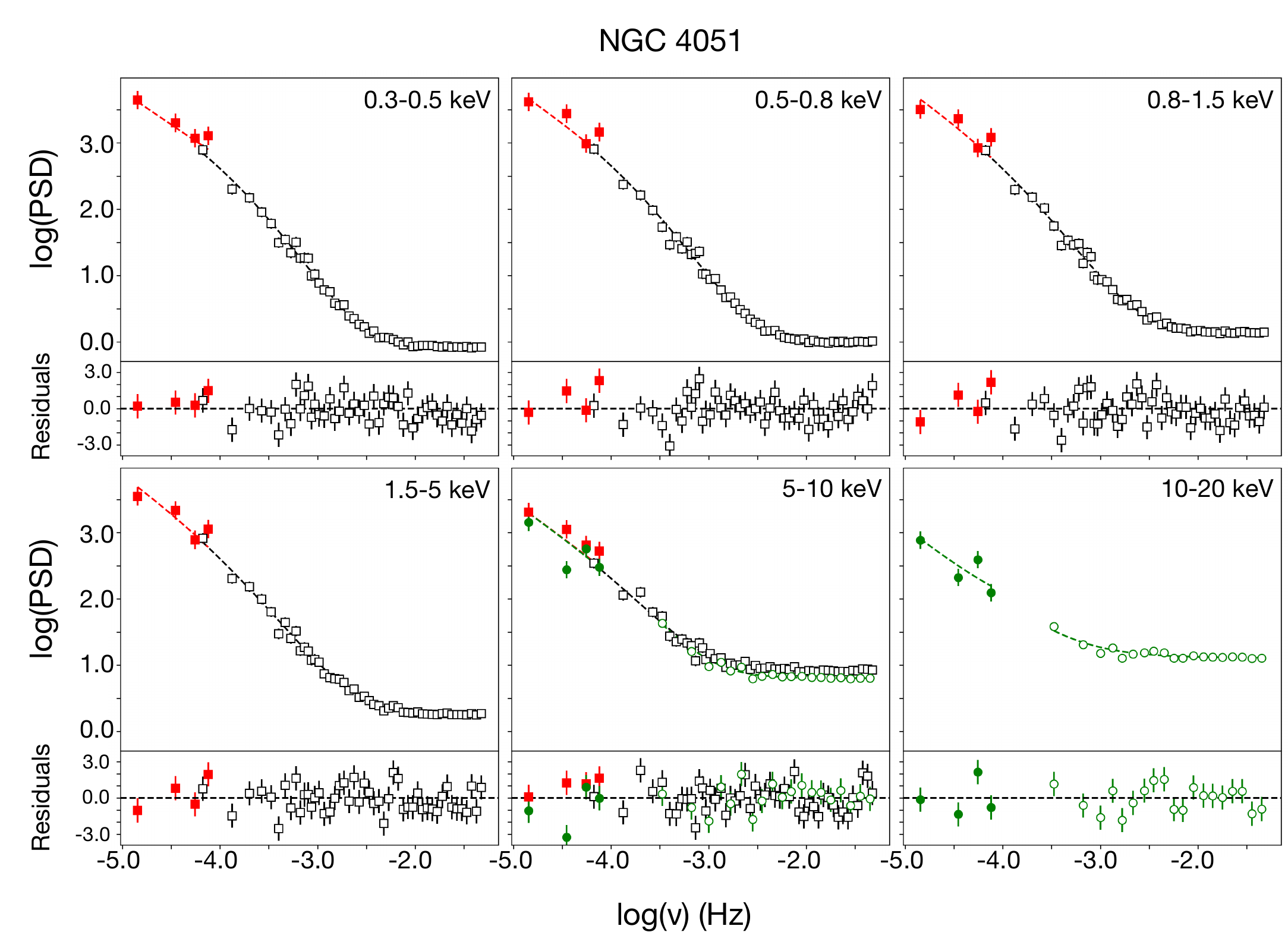}
  \caption{Observed PSDs of NGC 4051 in the various energy bands we considered in this work. Open black squares and open green circles indicate the  \xmm\ and \nustar\ high frequency power spectra, respectively. Errors are plotted in all points, but they are smaller than the symbol size at high frequencies. Filled red squares and filled green circles indicate the low-frequency \suzaku\ and \nustar\ power spectra, respectively. The dashed lines indicate the best-fit BPL models to the PSDs at energies lower than 10 keV. For the 10--20 keV band, the dashed green lines correspond to the best-fit PL model. The bottom sub-panels display the best-fit residuals, with points colour-coded in the same manner (residuals were calculated as the difference between observed and model PSDs divided by the error of the PSDs).}
  \label{fig:4051-PSD}
\end{figure*}


\begin{figure*}[htbp]
  \centering
  \includegraphics[width=0.9\textwidth]{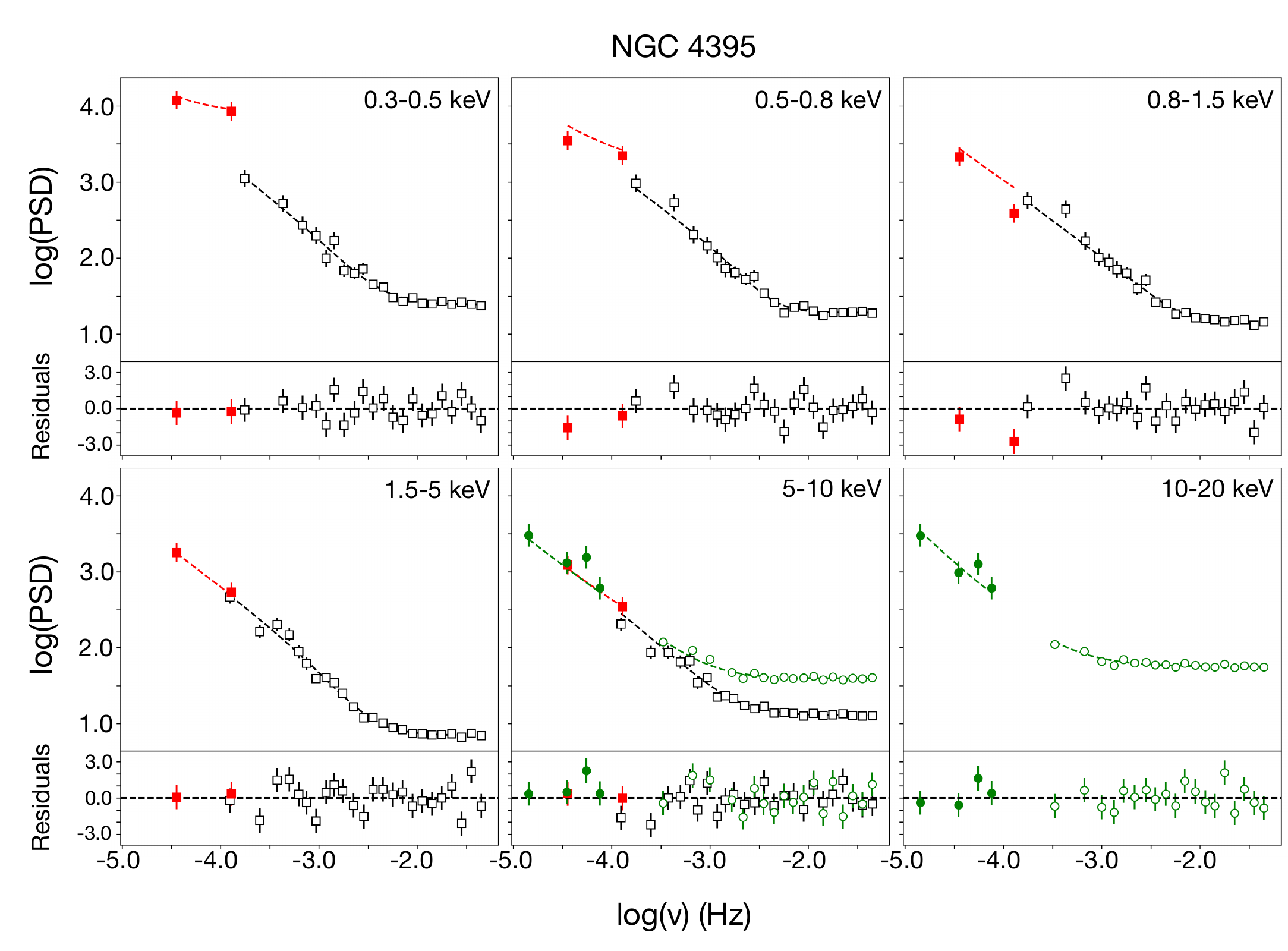}
  \caption{Same as in Fig.\,\ref{fig:4051-PSD}, but for NGC 4395.}

  \label{fig:4395-PSD}
\end{figure*}

The duration of the \xmm\ light curve segments, $T_{seg}$, was 15 and 8 ks for NGC 4051 and NGC 4395, respectively. The choice of $T_{seg}$ is a compromise between the need to study the power spectrum at low frequencies (which implies large $T_{seg}$), and the need for as many segments as possible to reduce $\sigma_{PSD_{final}}(f_j)$ and bring the distribution of log$[{\rm PSD}_{final}(f_j)]$ close to Gaussian.

It is also necessary that $T_{seg}$ is greater than the bending timescale, $T_b=1/\nu_b$.  If the PSD slope below $\nu_b$ is $\sim -1$, then the red-noise leakage will be minimal, provided that $T_{seg}>T_b$. In the opposite case, red-noise leakage can bias the observed PSD, since the high-frequency PSD slope is usually steeper than $-2$. The chosen $T_{seg}$ is larger than $T_b$ for both sources. 

There are $n_{seg}=50$ and 59 such segments for NGC 4051 and NGC 4395, respectively. However, there are issues with many of the NGC 4395 segments in the low-energy bands. This is a low-luminosity Seyfert, which is affected by absorption, both by neutral and ionised material \citep[see e.g.][]{Kammoun19}. Figure\,\ref{fig:countrates} shows a plot of the mean count rate of the 59 segments of the NGC 4395 light curves in the 0.3--0.5 and 1.5--5 keV bands versus the mean count rate in the 5--10 keV band (top and bottom panels, respectively). Variations in the 1.5--5 and 5--10 keV bands follow each other well, indicating that they are mainly due to intrinsic source variations, unless absorption variations affect the 5--10 keV band as well, which does not seem to be the case (see, for example, Fig. 5 in \citealt{Kammoun19}). However,  the top panel of Fig.\,\ref{fig:countrates} shows that the mean count rate in the 0.3--0.5 keV band is very low and does not appear to vary as the 5--10 keV band flux does. This is probably due to heavy absorption, which decreases the observed count rate below $\sim 2$ keV \citep{Kammoun19}. There is an exception, and this is the 2003 \xmm\ observation (ObsID 0142830101). The points in the top panel of Fig.\,\ref{fig:countrates} showing significant variability in the 0.3--0.5 keV band, which is correlated with the variability in the 5--10 keV band, are from this observation. 

We calculated the PSD at energies below 1.5 keV using segments from the 2003 \xmm\ observation only (12 segments) and segments from all observations. We cannot detect the intrinsic power spectrum (especially at high frequencies) above the Poisson noise level in the latter case. This is due to the low count rate when the source is heavily absorbed. The amplitude of the intrinsic variations in this case is comparable and/or smaller than the amplitude of the variations due to the experimental Poisson noise. We therefore  decided to work with the PSD which is calculated using segments from the 2003 observation only. The number of segments is smaller than 20 in this case. Hence, we calculated the mean of the log-periodogram estimates at two neighbouring frequencies. In this way, we average 24 log-periodogram estimates, and the resulting log(PSD) should be approximately Gaussian distributed \cite[e.g.][]{papadakis-lawrence1993}. 

To calculate the PSD in the two highest-energy bins, we used the 42 segments with a mean count rate greater than 0.15 counts/sec in the 5--10 keV band (indicated by the vertical dashed line in the left panel in Fig.\,\ref{fig:countrates}). As with the PSD estimation at low energies, if we use segments with smaller count rate, then we increase the average Poisson level, and the high-frequency PSDs are less well defined.  

We used the 10-sec binned \nustar\ light curves to calculate the high-frequency PSD in the 10--20 keV band for both sources. We created 102 and 114 segments that were 3000 s long, binned at $\Delta t=10$ s, for NGC 4051 and NGC 4395, respectively. In this case, $T_{seg}$ is determined by the presence of data gaps in the light curves due to Earth occultation. If we increase $T_{seg}$, then there will be long gaps in the light curves. We followed the same procedure as described above for the \xmm\ light curves and calculated the average log-periodogram estimates for each of the Fourier frequencies. 

\subsection{Estimation of the low-frequency PSD}

We used the 2900 s binned \suzaku\ and \nustar\ light curves to compute the PSD at low frequencies. Given the orbital period of these satellites, these light curves consist of pairs of points with a high and a low fraction of the exposure time. We kept the first point with the largest exposure fraction and then every other point in the light curve. In this way, the light curves consist of points that are binned over 2900 s, while $\Delta t =5800$ s. 

We note that in the case of the 2900 s binned \nustar\ light curves, the fraction of the exposure time is larger than 0.7 \footnote{This is the fraction of the exposure time in the \xmm\, SWM light curves} in more than 87\% of the points (and it is higher than 0.65 in the remaining points). However, this is not the case for the \suzaku\ light curves. The fractional exposure of all the points in the NGC 4051, 2900 s binned \suzaku\ light curves and in the single Suzaku observation of NGC 4395 is larger than 0.4 and 0.2, respectively. Fractional exposure is larger than 0.7 in 52\% and 32\% of the points in the NGC 4051 and NGC 4395 light curves, respectively. In the case of bins with low fraction of exposure the mean time of the observations may be different from the mean time of the bin by $\sim 1000-1200$ s, at most. By moving the mean time of the observations by that amount (in light curves which have duration of a day or several days), should affect the estimation of the PSD at frequencies higher than $\sim 10^{-3}$ Hz. However, we use the \suzaku\ light curves to estimate the PSD at frequencies lower than $\sim 8.6\times 10^{-4}$ Hz. An indication that the relatively low fractional exposure for many points in the \suzaku\ light curve does not significantly affect the estimation of the low-frequency PSD is provided by the comparison between the low-frequency \suzaku\ and \nustar\ PSDs in the  5--10 keV band. The two PSDs are almost identical in both sources (see the red filled squares and green filled circles in the 5--10 keV panels in Figs.\,\ref{fig:4051-PSD} and \ref{fig:4395-PSD}).

In the case of NGC 4051, we divided the \suzaku\ and \nustar\ light curves into eight and nine segments, respectively, with $T_{seg}=95.7$ ks. There are 16 points in each segment, and we computed the respective log-periodograms in eight frequencies. As before with the high-frequency PSD estimation of NGC 4395 at low energies, since $n_{seg}$ is smaller than 20, we calculated the mean of the log-periodogram estimates at two neighbouring frequencies. In this way, we average 16 and 18 log-periodogram estimates, and the resulting log(PSD) should be approximately Gaussian distributed. Consequently, the low-frequency part of the NGC 4051 PSDs is calculated at four frequencies, both for the \suzaku\ and the \nustar\ light curves. 

In the case of NGC 4395, there is only one \suzaku\ light curve. It is 232 ks long, with $N=80$ points. In this case, we calculated the log-periodogram in 40 frequencies, and we binned consecutive log-periodogram estimates over bins of size $m=20$. As long as the intrinsic power spectrum has a PL-like shape, binning in the log space does not introduce any bias. Consequently, there are two points in the \suzaku\, low-frequency part of the NGC 4395 power spectrum. Like NGC 4051, we divided the \nustar\ light curves of NGC 4395 into seven segments with $T_{seg}=95.7$ ks, and we calculated the average PSD in the same way as described above, so there are four points in the low-frequency \nustar\ PSDs in the higher energy bands. 


\begin{table*}[h!]
\caption{PSD best-fit Results}
\label{table:fit_params_combined}
\centering
\small 
\renewcommand{\arraystretch}{1.3}
\begin{tabular}{c|ccccc|ccccc}
\multicolumn{1}{c}{} & \multicolumn{5}{c}{\textbf{NGC 4051}} & \multicolumn{5}{c}{\textbf{NGC 4395}} \\
 & $\chi^2/\mathrm{dof}$ & $\chi^2/\mathrm{dof}$ & PSD$_{amp}$ & $\nu_b$  & $a_{\rm high}$ & $\chi^2/\mathrm{dof}$ & $\chi^2/\mathrm{dof}$ & PSD$_{amp}$ & $\nu_b$  & $a_{\rm high}$ \\
(keV) & (PL) & (BPL) & ($\times10^{-2}$) & ($\times10^{-3}$ Hz) & & (PL) & (BPL) & ($\times10^{-2}$) & ($\times10^{-3}$ Hz) & \\
\hline
0.3--0.5 & 105.2/50 & 49.2/49 & $3.0^{+0.5}_{-0.4}$ & $0.20^{+0.06}_{-0.04}$ & $2.16\pm0.06$ & 22.4/20 & 15.9/19 & $10.3^{+2.9}_{-2.3}$ & $2.15^{+1.44}_{-0.86}$ & $2.22^{+0.52}_{-0.42}$ \\
0.5--0.8 & 103.5/50 & 56.4/49 & $3.5^{+0.6}_{-0.5}$ & $0.17^{+0.05}_{-0.04}$ & $2.09\pm0.06$ & 41.8/20 & 20.4/19 & $7.1^{+1.2}_{-1.1}$ & $2.53^{+0.74}_{-0.57}$ & $2.88^{+0.68}_{-0.55}$ \\
0.8--1.5 & 93.7/50 & 62.6/49 & $3.6^{+0.8}_{-0.7}$ & $0.13^{+0.06}_{-0.04}$ & $1.98\pm0.06$ & 42.1/20 & 27.3/19 & $4.7\pm0.6$ & $4.34^{+1.14}_{-0.90}$ & $2.82^{+0.77}_{-0.60}$ \\
1.5--5 & 76.3/50 & 57.7/49 & $4.2^{+1.2}_{-1.0}$ & $0.09^{+0.06}_{-0.04}$ & $1.88\pm0.06$ & 58.5/23 & 29.2/22 & $3.1\pm0.4$ & $1.56^{+0.43}_{-0.34}$ & $2.32^{+0.24}_{-0.22}$ \\
5--10 & 110.3/73 & 93.5/72 & $1.5\pm0.3$ & $0.19^{+0.12}_{-0.08}$ & $1.88^{+0.14}_{-0.13}$ & 54.4/44 & 48.0/43 & $1.9\pm0.3$ & $1.04^{+0.54}_{-0.36}$ & $2.01^{+0.31}_{-0.27 }$\\
10--20 & 26.0/21 &  & $<1$ & $>0.12$  & $1.8$\tablefootmark{a}  & 17.7/20 &  & $<7.2$ & $>0.03$  & $2$\tablefootmark{a} \\
\hline
\end{tabular} \\
\tablefoot{PSD best-fit results for NGC 4051 and NGC 4395 in the various energy bands. Columns (2,3), and (7,8) list the best-fit $\chi^2$ for the PL and BPL models to the NGC 4051 and NGC 4395 PSDs, respectively. Columns (4,5,6) and (9, 10, 11) list the best-fit BPL parameters. Errors correspond to 1$\sigma$ confidence range.}
\parbox{0.96\textwidth}{\footnotesize
\raisebox{-0.3ex}{\tablefootmark{a}}~The BPL high-frequency slope is frozen at -1.8 and -2.0 for NGC 4051 and NGC 4395, respectively.
}

\end{table*}

\section{Model fits to the PSDs}\label{modelfits}

The resulting PSDs are shown in Figs.\,\ref{fig:4051-PSD} and \ref{fig:4395-PSD} for NGC 4051 and NGC 4395, respectively. The high-frequency data in the PSDs are further binned to reduce the errors of the log-periodogram estimates. 

We fitted the PSDs in all energy bands with two models. The first one is a PL model defined as follows: 
\begin{equation}\label{eq:PL}
    \log[{\rm PSD}_{PL}(\nu)]=\log \left[ A \left( \frac{\nu}{\nu_0} \right)^{-a} + C_{PN} \right],
\end{equation}
where A is the PL amplitude (equal to the power spectrum at $\nu=\nu_0=3\times 10^{-4}$ Hz), and $a$ is the slope of the PL. The second model is a BPL model which is defined as follows: 

\begin{equation}\label{eq:BPL}
    \log[{\rm PSD}_{BPL}(\nu)]=\log\left[ \frac{A \ \nu^{-1}}{ 1 + \left( \frac{\nu}{\nu_b} \right)^{\alpha_{high}-1}  } + C_{PN}\right]. 
\end{equation}

\noindent Defining the BPL model in this way implies that the slope of the power spectrum is $-1$ at low frequencies, up to $\nu_b$, where the slope increases to $\alpha_{high}$. Regarding the normalisation of the model, $A/2$ is equal to PSD$_{BPL}(\nu_b)\times \nu_b$. This is the value that we list in the table with our results (and we call it PSD$_{amp}$). 

The constant $C_{PN}$ in Eqs. (\ref{eq:PL}) and (\ref{eq:BPL}) is the Poisson noise power level, which is different for the PSDs that we calculated using the \xmm, \suzaku, and \nustar\ light curves. We let this constant vary freely during the model fitting for the high-frequency PSDs, as $C_{PN}$ is well defined in these cases. However, this is not the case for the low-frequency \suzaku\ and \nustar\ PSDs. The Poisson noise constant is not well sampled in these PSDs. This noise component can be predicted by the error of the points in the light curves. The PN level is given by the following equation:

\begin{equation}\label{eq:poisson_noise}
    C_{PN}= \frac{2 \Delta t \ \overline{\sigma_{err}^2}}{\overline{x}^2},
\end{equation}

\noindent where $\Delta t$ is the size of the light curve bin, $\overline{\sigma_{err}^2}$ is the mean squared error of the points in the segment, and $\overline{x}^2$ the square of the mean count rate. The constant Poisson noise for the fits to the low-frequency \suzaku\ and \nustar\ PSDs was determined by computing the mean log$(C_{PN})$ of all segments of the light curve, using Eq.\,\ref{eq:poisson_noise}. 

The PSD best-fit results for both models and both sources are listed in Table \ref{table:fit_params_combined}. In general, the BPL model produces lower $\chi^2$ values. The PL model does not fit well (i.e. $p_{null}<0.01$) any of the PSDs in the 0.3-10 keV bands in NGC 4051. In contrast, the BPL model fits all the PSDs very well. In NGC 4395, the PL model does not fit the PSDs well in the 0.5--0.8, 0.8--1.5 and 1.5--5 keV bands. It provides acceptable fits to the PSDs in the 0.3--0.5 and 5--10 keV bands, but even in these two bands the $F-$test indicates that the BPL fits the PSDs significantly better ($\Delta\chi^2=12.9$ for 2 dof, $F$ statistic = 6.26, and $p_{null}=0.003$). We therefore conclude that we have detected the PSD bending frequency in all PSDs of both sources in the 0.3--10 keV band. Dashed lines in Figs.\,\ref{fig:4051-PSD} and \ref{fig:4395-PSD} show the best-fit BPL models to the high- and low-frequency PSDs at energies lower than 10 keV. The PL model fits well the 10--20 keV band PSDs (the dashed lines in the 10--20 keV PSDs in Figs.\,\ref{fig:4051-PSD} and \ref{fig:4395-PSD} show the PL best-fit to the PSDs). 

The best-fit slopes are all consistent with -1. This result indicates that we cannot detect the bending frequency because we cannot probe the high-frequency part of the PSDs in these energy bands. This is because the Poisson noise level is quite high in these PSDs (and the PSD amplitude is rather low, at least in NGC 4051, see, e.g. the bottom right panel in Fig.\,\ref{fig:4051-PSD}). 

To place some constraints on $\nu_b$, we fitted the respective PSDs with a BPL model keeping $\alpha_{high}$ fixed at 1.8 and 2 in NGC 4051 and NGC 4395, respectively (these slopes are roughly equal to the best-fit $\alpha_{high}$ values in the 5--10 keV band). The corresponding upper and lower 99\% limits in the 10--20keV PSD$_{amp}$ and $\nu_b$ are listed in Table \ref{table:fit_params_combined}. 


\begin{figure*}[htbp]
  \centering
  \includegraphics[width=0.95\textwidth]{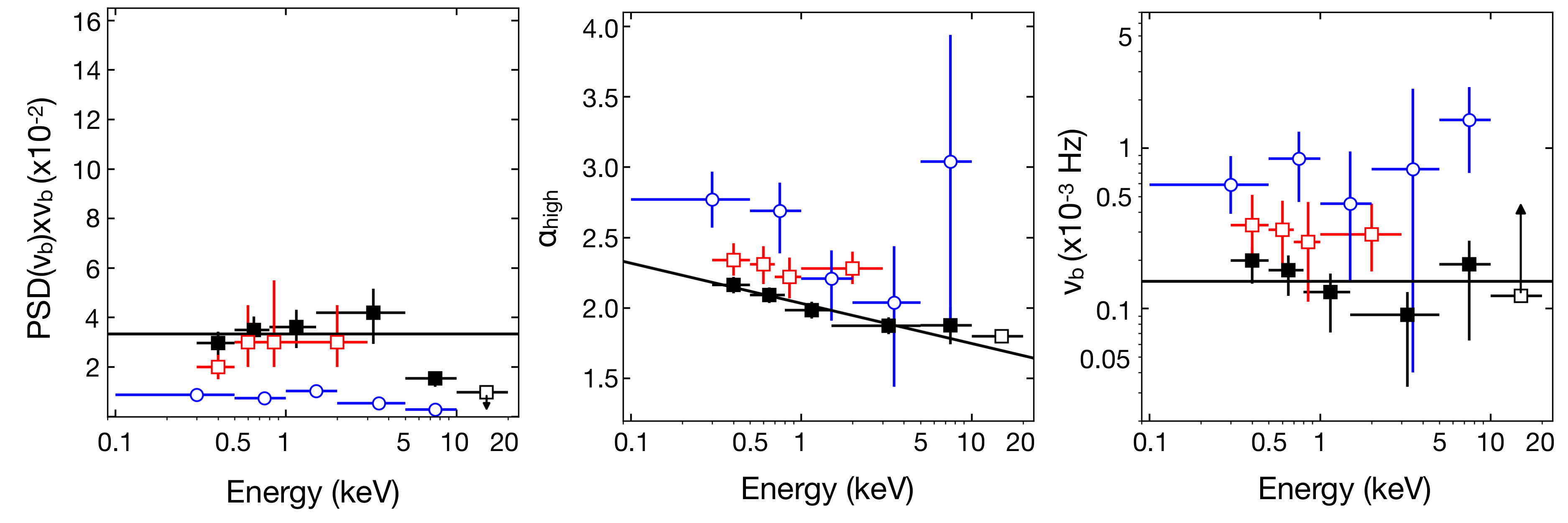}
  \caption{The energy dependence of the PSD parameters of NGC 4051. The solid black squares show the best-fit, BPL model parameters plotted as a function of energy for NGC 4051. Open red squares and open blue circles show the respective results from \citet{rani-etal2025}, and from \citet{mchardy-etal2004}, respectively. The solid lines indicate the weighted PSD$_{amp}$ of the four lowest energy points in the left panel, the best-fit log-linear model to the $\alpha_{high}-E$ data in the middle panel, and the weighted bending frequency in the right panel. Note that, for comparison reasons, the y axis limits are the same in the relevant panels in this figure and in Fig.\,\ref{fig:4395-fit-params}.}
  \label{fig:4051-fit-params}
\end{figure*}

\section{The energy dependence of the power spectrum}\label{energy-dep}

The black squares in Fig.\,\ref{fig:4051-fit-params} show the best-fit parameters of the BPL model plotted as a function of energy in NGC 4051: PSD amplitude, high-frequency slope, and bending frequency versus energy in the left, centre, and right panels, respectively. The blue circles and red squares indicate the best-fit parameters of \cite{mchardy-etal2004} and \cite{rani-etal2025}, respectively. The \cite{mchardy-etal2004} results are based on the PSD modelling of a single \xmm\ light curve. They also used the BPL model to fit the PSDs and kept the low-frequency PSD slope fixed at 1.1 (i.e. similar to our approach). They calculated the PSD in similar energy bands, and the trend of the best-fit parameters with energy is in agreement with ours. However, the error of the best-fit parameters from our PSD modelling is smaller, because we use all of the archival \xmm\ observations, as well as longer light curves (\citealt{mchardy-etal2004} report the PSD amplitude without errors). The fact that we determine the PSD at lower (and higher) frequencies with smaller errors may also explain the systematic difference between \cite{mchardy-etal2004} and our results. We detect the PSD bending frequency at lower frequencies, which explains why the PSD amplitude turns out to be larger in our case. At the same time, there is a wider frequency range at high frequencies to determine $\alpha_{high}$ in our case. It is interesting that $\alpha_{high}$ appears to be quite flatter compared to the \cite{mchardy-etal2004} results. \cite{rani-etal2025} performed a PSD analysis of the NGC 4051 light curves in a narrower energy range. They used the same \xmm\ data sets, but the addition of the \suzaku\ and \nustar\ long light curves allowed us to fit the PSD over a wider frequency range. For the same reasons that we discussed above, since our best fit $\nu_b$ is lower than the \cite{rani-etal2025} results, the best-fit PSD$_{amp}$ is also larger. 

The left panel in Fig.\ref{fig:4051-fit-params} suggests that PSD$_{amp}$ may increase from 0.4 to 3 keV and then decrease considerably at higher energies. The upper limit at 15 keV indicates that the decrease in the PSD amplitude continues at energies higher than 10 keV. The apparent increase in the PSD amplitude with the energy below 3 keV is not significant. The weighted mean of the PSD amplitude in the first four energy bands is equal to $\overline{\rm PSD}_{amp}=3.3(\pm0.3)\times 10^{-2}$ (indicated by the horizontal solid line in the left panel of Fig.\,\ref{fig:4051-fit-params}). The amplitude of the power spectrum in these bands is consistent with $\overline{\rm PSD}_{\rm amp}$ (within errors). The middle panel of Fig.\,\ref{fig:4051-fit-params} shows that $\alpha_{high}$ flattens with increasing energy. A log-linear relation of the form $\alpha_{high}(E) = A + B \log(E)$ fits the data well (\,$\chi^2=1/3$ dof), with $A=2.03\pm0.03$ and $B=0.29\pm0.08$. 

Finally, we find no evidence of a significant correlation between $\nu_b$ and energy in NGC 4051 (right panel in Fig.\,\ref{fig:4051-fit-params}). The weighted mean of the bending frequency, $\bar\nu_{b,{\rm NGC 4051}}$, is equal to $1.5(\pm0.2)\times 10^{-4}$ Hz. The horizontal solid line in this panel indicates the mean bending frequency. This line fits the data very well ($\chi^2$ = 3.2/4 dof, $p_{null}=0.52$). This result indicates that the bending frequency in the X-ray PSD of NGC 4051 does not depend on energy. 

\begin{figure*}[htbp]
  \centering
  \includegraphics[width=0.95\textwidth]{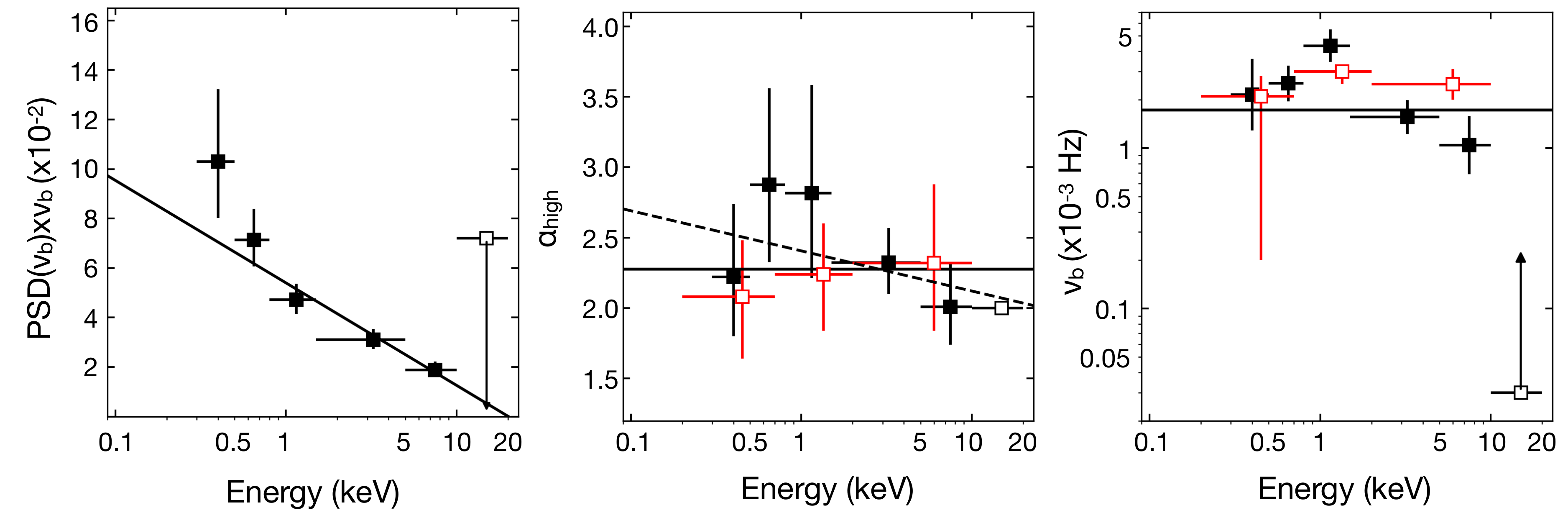}
  \caption{Same as in Fig.\,\ref{fig:4051-fit-params}, but for NGC 4395. Open red squares in this case indicate the results of \cite{vaughan-etal2005}. The solid lines indicate the best-fit log-linear model to the PSD$_{amp}$ vs $E$ plot (left panel), the weighted mean $\alpha_{high}$ and $\nu_b$ (middle and right panels, respectively). The dashed line in the middle panel indicate the best-fit log-liner model with a slope equal to the slope of the bst-fit to the $\alpha_{high}-E$ data in NGC 4051. } 

  \label{fig:4395-fit-params}
\end{figure*}

The results on the energy dependence of the BPL parameters of NGC 4395 are shown in Fig.\,\ref{fig:4395-fit-params}. Our results agree well with the results of \cite{vaughan-etal2005} (shown with the open red squares). The PSD amplitude decreases strongly with energy. The solid black line in the left panel of Fig.\,\ref{fig:4395-fit-params} shows the best-fit log-linear function of the form PSD$_{amp}(E) = A + B \log(E)$. It fits the data well (\,$\chi^2=3.1/$3 dof), with $A=5.4(\pm0.5)\times 10^{-2}$ and $B=0.04\pm0.01$). The solid line in the middle panel indicates the weighted mean of $\alpha_{high}$ (which is $2.28 \pm 0.16$). This line fits the data well (\,$\chi^2=2.5/$4 dof). This result indicates that the PSD high-frequency slope does not depend on the energy in this source. However, this result may depend on the fact that the errors of $\alpha_{high}$ are larger in NGC 4395 (which is not surprising, given the fact that the bending frequency is much higher in this source). The dashed line in the middle panel of Fig.\,\ref{fig:4395-fit-params} shows the best fit of a log-linear line with a slope equal to the best-fit slope of the solid line plotted in the middle panel of Fig.\,\ref{fig:4051-fit-params} in the case of NGC 4051. The fit is good (\,$\chi^2=1.6/$4 dof); although this result is not conclusive, it shows that it is possible that $\alpha_{high}$ flattens with increasing frequency in NGC 4395 like in NGC 4051.

The solid line in the right panel of Fig.\,\ref{fig:4395-fit-params} shows the weighted mean of the bending frequency in the X-ray PSD of NGC 4395 at all frequencies ($\bar{\nu}_b=1.73(\pm0.25)\times 10^{-3}$ Hz.). This line fits the data with $\chi^2=10.8/$4 dof. The null hypothesis probability is 0.03, which is higher than the limit of 0.01 that is usually accepted to assess whether a fit is statistically accepted or not. We note that $p_{null}=0.03$ corresponds to a $\sim 1.9\sigma$ effect, which is hardly accepted as significant. Nevertheless, to investigate the possibility of a non-constant bending frequency, we fit the data with a straight line in the log-log space, and find a best-fit slope of $-0.40\pm0.16$. The best-fit improves by $\Delta\chi^2=3.1$  for one extra parameter, which is not statistically significant (according to the F-test). We therefore conclude that the hypothesis of a constant bending frequency is fully consistent with the data. There is an indication that the bending frequency may decrease with increasing energy, but this is not statistically significant.


\section{Discussion}\label{disc}

In this study, we used data from \xmm, \suzaku, and \nustar\ to estimate the PSD of NGC 4051 and NGC 4395 in five energy bands within the range of 0.3--10 keV. We also considered the 10--15 and 15--20 keV band \nustar\ light curves but, because of the low count rate, we finally studied the combined 10--20 keV band light curves only. 

We used data from \xmm\ and \nustar\ to probe the high-frequency PSD, and data from \suzaku\ and \nustar\ to calculate the power spectrum at frequencies lower than the range covered by \xmm. This resulted in PSDs that extended above the Poisson noise level from $\sim10^{-5}$ Hz up to 2 to 3 orders of magnitude in the frequency range, depending on the energy band. This is the first time power spectra have been calculated for both sources in so many energy bands, over a broad range of frequencies, with high accuracy. Our results indicate interesting relations between the PSD parameters and energy, which can put constraints on various theoretical models, as we discuss below. 

\subsection{The energy dependence of the PSD bending frequency.} 

Our results indicate that the bending frequency remains constant with the energy in both sources. These results are not in agreement with the propagating fluctuations model (\citealt{lyubarskii1997}; \citealt{kotov-etal2001}; \citealt{arevalo-uttley2006}). According to this model, the corona accretes, and its temperature increases as the radius decreases. This causes higher-energy photons to originate closer to the BH. Accretion rate fluctuations occur at each radius on a characteristic timescale which is equal to the local viscous timescale. This timescale decreases inwards, i.e. as the radius decreases. Consequently, since higher-energy photons are emitted from the inner corona,  this model predicts an increase in the bending frequency with increasing energy. For example, Figure 2 in \cite{arevalo-uttley2006} shows that the bending frequency of the 'hard' energy band power spectrum (indicated with the dashed line in Figure 2) is significantly higher than $\nu_b$ of the 'soft' band PSD (dotted line). 

It is possible that the X-ray corona is located within the disc and that accretion fluctuations do not propagate within the X-ray corona. The observed X-ray variations could be caused by accretion-rate variations in the accretion disc, which propagate inwards, down to the inner disc radius (which coincides with the outer radius of the corona). In this case, the bending frequency in the PSD should be the same in all energy bands, and it would be equal to the viscous timescale at the inner disc radius. However, accretion rate fluctuations propagating through the corona are necessary to explain the continuous, hard time-lags that are observed in Seyfert galaxies, like NGC 4051 for example \citep{epitropakis}. Therefore, we should expect an increase in $\nu_b$ with increasing energy, which is not what we observe. 

Another possibility that could be consistent with our data is a highly magnetised corona. According to \citet{haardt-etal1994}, the energy that powers the corona could be stored in magnetic field structures above the accretion disc \citep[see also][]{galeev}. These structures accumulate energy over a characteristic 'charging' timescale. This energy is subsequently released over a 'discharge' timescale. This process heats up electrons within the corona, which in turn produce X-rays via inverse Compton scattering of optical and ultraviolet photons emitted from the disc. In essence, X-ray variability is governed by this cycle, which is naturally independent of energy, and the bending frequency could correspond to one of the two timescales that govern this cycle's time evolution. According to \citet{paolillo-papadakis2025}, the charging timescale could be responsible for the bending timescale detected in Seyferts so far (see their Figure 27). Therefore, this possibility could explain the fact that $\nu_b$ does not depend on the energy in NGC 4051 and NGC 4395. 

We note that the average bending frequency in the X-ray PSD of NGC 4395 is $\sim 10$ times higher than $\nu_b$ in NGC 4051. If characteristic timescales increase proportionally to the BH mass in accreting systems, then we would expect the BH mass in NGC 4395 to be $\sim 0.1$ of the BH mass in NGC 4051. Therefore, our results favour BH mass estimates that are smaller/larger than the values we mentioned in Sect.\,\ref{intro} for NGC 4395 and NGC 4051, respectively. For example, \citet{edri-etal2012}, found a BH mass of 5$(\pm 2.6)\times 10^4$ M$_{\odot}$ for NGC 4395. This value would imply a ratio of 0.06$\pm 0.03$ for the BH mass estimates for the two sources, in agreement with the ratio of the respective bending timescales we measure in this work.

\subsection{The energy dependence of the high-frequency PSD slope.}

Our results show that the high-frequency slope becomes flatter with increasing energy in NGC 4051. The best-fit slope of the line plotted in the middle panel of Fig.\,\ref{fig:4051-fit-params} is $-$0.29$\pm 0.08$, which implies that the flattening of the PSD slope with energy is significant at the 3.6$\sigma$ level in this object. This trend between $\alpha_{high}$ and energy is not directly visible for NGC 4395, but may be present in this source as well. The dashed line plotted in the middle panel of Fig.\,\ref{fig:4395-fit-params} indicates that the same flattening of the high-frequency PSD slope with increasing energy that we see in NGC 4051 may also be true for NGC 4395. 

The dependence of the PSD slope on energy can put important constraints on the physical properties of the X-ray corona in AGN. \citet{papadakis-lawrence1995} were the first to notice this trend in their study of the NGC 4051 power spectrum with the use of {\it EXOSAT} light curves. They also pointed out the importance of this effect. If X-rays are produced via inverse Compton scattering in a corona with uniform temperature, then the PSD is expected to (significantly) steepen at higher energies. The reason is that the increased number of scattering events required to produce higher-energy photons eliminates fast variations, and therefore results in lower variability at higher frequencies. Therefore, one would expect the high-frequency PSD slope to increase significantly (i.e.to steepen) with increasing energy. This is opposite to what we observe in NGC 4051, and potentially to NGC 4395 as well. 

On the other hand, the data indicate that the PSD slope in NGC 4395 may be constant with energy. The same may also be the case in NGC 4051 at low energies. The best-fit values in Table \ref{table:fit_params_combined} show that the high-frequency slope is roughly constant above $\sim 1.5$ keV, and steepens at lower energies, where the soft-excess component is present. If this component is due to X-ray reflection, then we expect it to be less variable at high frequencies (as the disc reprocessing will smooth out the fast variations), which could lead to a steepening of the PSD slope. However, even if the PSD slope remains constant with energy, this is still against the prediction of thermal Comptonisation. Our results regarding the high-frequency PSD strongly suggest that the X--ray corona in AGN cannot be just a region with uniform energy. It must be dynamic, and either it is a single medium with a non-uniform temperature or it is composed of multiple emitting regions with different temperatures. 

\subsection{The energy dependence of the PSD amplitude.}
The energy dependence of the PSD amplitude remains an open question. As \citet{paolillo-papadakis2025} reports, the amplitude of the PSD in Seyfert galaxies remains roughly constant at frequencies below $\nu_b$. In the $PSD(\nu)\times \nu$ space, the low-frequency PSD amplitude in the 2--10 keV band is approximately $\sim$0.01--0.02.  Furthermore, \citet{papadakis-binas2024} found that, for AGNs, $\mathrm{PSD}(\nu)\times\nu$ is $\sim 0.014$ in the 14--195 keV band (at low frequencies). This result suggests little to no energy dependence on the average PSD amplitude of AGNs over a broad energy range. 

Our results show that the PSD amplitude decreases with increasing energy at energies higher than 1.5 keV in both sources. One possibility is that the continuum variability at these energy bands is diluted by the less variable reflection component (from distant or nearby material). In this case, the contribution from the  less variable reflection component should increase steadily with increasing energy, but this possibility must be examined in detail to investigate whether it can indeed explain our results.

At energies below 1.5 keV, the PSD amplitude may decrease (slightly) with decreasing energy in NGC 4051 (see the left panel in Fig.\,\ref{fig:4051-fit-params}). This could be due to the presence of a soft excess, which is believed to be less variable than the continuum X-ray variability. However, the PSD amplitude continues to increase with decreasing energy in the case of NGC 4395. An extra  soft component (on top of the continuum PL-like spectrum) has been identified  by \cite{Kammoun19} in this source. They modelled it as emission from collisionally ionised diffuse gas, and they argued that this is not the typical soft-excess component that is detected in many other AGNs. Irrespective of its origin, our results indicate that this component cannot be dominant at low energies in this source. If that were the case, we would expect the PSD amplitude to decrease with decreasing energy, which is not the case. Variable absorption could also explain the increase in PSD amplitude at low frequencies in NGC 4395. This is because absorption can decrease the observed mean at low energies and can also increase the variance (if absorption variations operate on timescales shorter than a day). This may be the case at energies below 0.8 keV, where the PSD amplitude appears to increase even more than the extrapolation of the PSD amplitude versus energy relation at higher energies. However, this is not a significant effect.

If the decrease in PSD amplitude with increasing energy is a property of the intrinsic variability mechanism, then it is probably not consistent with the fluctuating accretion rate model. Figure 2 in \cite{arevalo-etal2006}, for example, shows that the PSD amplitude below the frequency break should be the same in all energy bands. In fact, this figure shows that, if anything, the PSD amplitude should be smaller in the soft band (contrary to our results). Perhaps the decrease in the variability amplitude could be the result of inverse Comptonisation, since the higher number of scatterings for the higher energy photons should result in a decrease in the variability amplitude with increasing energy.

An additional possibility could be the number of individual sources that contribute to each energy band. As we argued in the previous section, the flattening of the PSD with increasing energy implies that there must be more than one X-ray emitting source, with different temperatures. The number of scatterings that are necessary for a photon to reach a given energy decreases with increasing temperature of the corona. Therefore, if the photons that are emitted in the 5--10 keV band are predominantly emitted by X-ray emitting regions with temperature larger than the temperature of the region that mainly emits the 0.3--0.5 keV photons, then the PSD slope could be flatter in that energy band. However, since inverse Comptonisation produces broad-band spectra, the number of the high-temperature coronae must be larger; hence, assuming simple Poisson statistics, we would expect the PSD amplitude to decrease with increasing energy. 

Some studies have proposed models in which X-rays are emitted by multiple 'active' regions. For example, \cite{galeev} proposed that X-rays in accreting objects are emitted by hot plasma loop structures that emerge from the inner disc. They did not consider variability, although \cite{haardt-etal1994} studied the basic variability properties of their model. \cite{poutanen99} proposed that X-rays could be produced in compact magnetic flares at small radii from the central BH. Similarly, \cite{merloni01} proposed a model in which magnetic flares shining above a standard accretion disc produce X-rays via inverse Compton scattering of soft photons (both intrinsic and reprocessed thermal emission from the accretion disc and locally produced synchrotron radiation). \cite{ghiselini04} proposed the idea of a 'failed jet' in radio-quiet AGNs. According to their model, blobs of material may be ejected, reaching a maximum radial distance, and then fall back, colliding with the blobs produced later and still moving outwards. These collisions dissipate the bulk kinetic energy of the blobs by heating the plasma, and could be the regions where X-rays are emitted from AGNs. 

Most of these models do not consider in detail the evolution of the PSD with energy, although there are exceptions. For example, the \cite{poutanen99} model predicts a similar high-frequency PSD slope at low and high energies (see their Fig. 2). Recently, \cite{zhang23} showed that a multiple of X-ray emitting sources (whose origin could be explained by the \cite{ghiselini04} model) can predict time lags that are similar to those observed in AGN, but they did not make any predictions about the power spectrum properties of their model. We plan to study the PSD predictions of this model in the near future and investigate whether it can be consistent with the energy dependence of the PSD parameters that we report. 

In summary, our study suggests that the dependence of the main parameters of the X-ray power spectrum in AGN, namely the PSD amplitude, high-frequency slope, and bending frequency, on energy, is complex and not necessarily strictly in line with the predictions of current models. This suggests that the X-ray corona is likely a more complex object than is currently thought. More studies are needed to determine the energy dependence of the power spectra in more Seyfert galaxies to understand the structure of the X-ray corona in these objects.  

     

%
%

\bibliographystyle{aa}      
\bibliography{bibliography}   

\begin{appendix}
\onecolumn
\section{Table of Observations}
\begin{table*}[h!]
\caption {The \xmm, \suzaku, and \nustar\ Observations}
\label{table:obs_log} 
\centering
\begin{tabular}{ccrccr}

\multicolumn{6}{c}{\bf XMM Observations}\\ 

\multicolumn{3}{c}{\bf NGC 4051} & \multicolumn{3}{c}{\bf NGC 4395}\\ \hline
\shortstack{Observation \\ ID} & \shortstack{Obs. Start \\ (yyyy-mm-dd hh:mm:ss)} & \shortstack{Duration \\ (sec)} & \shortstack{Observation \\ ID} & \shortstack{Obs. Start \\ (yyyy-mm-dd hh:mm:ss)} & \shortstack{Duration \\ (sec)}\\ \hline
    0109141401 & 2001-05-16 11:52:05 & 105000 & 0112521901 & 2002-05-31 00:38:06 & 8000 \\ 
    0157560101 & 2002-11-22 06:09:51 & 45000  & 0112522001 & 2002-06-12 18:17:26 & 8000\\ 
    0606320101 & 2009-05-03 10:28:32 & 45000  & 0142830101 & 2003-11-30 03:17:55 & 96000 \\
    0606320201 & 2009-05-05 10:21:56 & 30000  & 0744010101 & 2014-12-28 10:07:16 & 48000 \\
    0606320301 & 2009-05-09 10:07:38 & 30000  & 0824610101 & 2018-12-13 06:16:19 & 80000 \\ 
    0606320401 & 2009-05-11 10:00:35 & 30000  & 0824610201 & 2018-12-19 05:51:22 & 24000 \\
    0606321301 & 2009-05-15 13:16:11 & 30000  & 0824610301 & 2018-12-31 05:11:45 & 48000 \\ 
    0606321401 & 2009-05-17 09:41:17 & 30000  & 0824610401 & 2019-01-02 05:03:24 & 96000 \\ 
    0606321501 & 2009-05-19 09:34:48 & 30000  & 0913600101 & 2022-12-10 08:10:57 & 24000 \\
    0606321601 & 2009-05-21 09:27:52 & 30000  & 0913600601 & 2022-12-19 07:38:49 & 24000 \\ 
    0606321701 & 2009-05-27 10:50:16 & 30000  & 0913600701 & 2022-12-22 15:29:42 & 16000 \\
    0606321801 & 2009-05-29 09:15:27 & 30000  & & & \\
    0606321901 & 2009-06-02 11:04:29 & 30000  & & & \\
    0606322001 & 2009-06-04 10:41:05 & 30000  & & & \\
    0606322101 & 2009-06-08 08:40:24 & 30000  & & & \\
    0606322201 & 2009-06-10 08:21:42 & 30000  & & & \\
    0606322301 & 2009-06-16 08:29:06 & 30000  & & & \\
    0830430201 & 2018-11-07 10:34:26 & 75000  & & & \\
    0830430801 & 2018-11-09 09:48:58 & 75000  & & & \\
\hline
 & & & & & \\
 
\multicolumn{6}{c}{\bf Suzaku Observations} \\
\multicolumn{3}{c}{\bf NGC 4051} & \multicolumn{3}{c}{\bf NGC 4395}\\ \hline
\shortstack{Observation \\ ID} & \shortstack{Obs. Start \\ (yyyy-mm-dd hh:mm:ss)} & \shortstack{Duration \\ (sec)} & \shortstack{Observation \\ ID} & \shortstack{Obs. Start \\ (yyyy-mm-dd hh:mm:ss)} & \shortstack{Duration \\ (sec)}\\ \hline
  700004010 & 2005-11-10 19:14:14 & 223300 & 702001010 & 2007-06-02 14:30:03 &  232000 \\
  703023010 & 2008-11-06 07:39:03 & 495900 & & & \\
  700004020 & 2008-11-23 16:48:00 & 162400 & & & \\
\hline
 & & & & & \\
\multicolumn{6}{c}{\bf NuSTAR Observations} \\

\multicolumn{3}{c}{\bf NGC 4051} & \multicolumn{3}{c}{\bf NGC 4395}\\ \hline
\shortstack{Observation \\ ID} & \shortstack{Obs. Start \\ (yyyy-mm-dd hh:mm:ss)} & \shortstack{Duration \\ (sec)} & \shortstack{Observation \\ ID} & \shortstack{Obs. Start \\ (yyyy-mm-dd hh:mm:ss)} & \shortstack{Duration \\ (sec)}\\ \hline
  60001050002 & 2013-06-17 16:41:05 & 17400  & 60061322002 & 2013-05-10 02:31:03 & 40600\\
  60001050003 & 2013-06-17 21:21:10 & 87000  & 60802027002 & 2022-12-10 13:46:08 & 95700\\
  60001050006 & 2013-10-09 20:01:06 & 95700  & 60802027004 & 2022-12-14 11:01:06 & 89900\\
  60001050008 & 2013-10-09 20:01:06 & 104400 & 60802027006 & 2022-12-18 21:06:12 & 98600\\
  60401009002 & 2018-11-04 12:56:10 & 611900 & 60802027008 & 2022-12-22 10:26:07 & 98600\\
              &                     &        & 60802027010 & 2022-12-26 10:56:12 & 98600\\
              &                     &        & 60802027012 & 2022-12-30 09:46:13 & 107300\\
              &                     &        & 91001640002 & 2024-11-08 08:06:09 & 153700\\
\hline
\end{tabular}
\tablefoot{Table of the \xmm, \suzaku, and \nustar\ observations we used. In the third column we list the duration of the final light curves we used to compute the power spectrum. }
\end{table*}

\end{appendix}

\end{document}